\documentclass[5p,twocolumn]{elsarticle}

\usepackage{graphicx}
\usepackage{dcolumn}
\usepackage{bm}
\usepackage[colorlinks,allcolors=blue]{hyperref}

\begin{document}


\begin{frontmatter}
\title{$\omega$ structure in steel: a first-principles study}

\author[add1]{Yuji Ikeda}
\ead{ikeda.yuji.6m@kyoto-u.ac.jp}

\author[add1,add2,add3,add4]{Isao Tanaka}
\ead{tanaka@cms.mtl.kyoto-u.ac.jp}
\address[add1]{Center for Elements Strategy Initiative for Structure Materials (ESISM), Kyoto University, Kyoto 606-8501, Japan}
\address[add2]{Department of Materials Science and Engineering, Kyoto University, Kyoto 606-8501, Japan}
\address[add3]{Center for Materials Research by Information Integration, National Institute for Materials Science (NIMS), Tsukuba 305-0047, Japan}
\address[add4]{Nanostructures Research Laboratory, Japan Fine Ceramics Center, Nagoya 456-8587, Japan}

\date{\today}

\begin{abstract}
Recent experimental works reported observation of the $\omega$ structure in steel.
Here, stability of the $\omega$ structure in steel is investigated
based on first-principles with special interests in effects of interstitial C atoms.
The interstitial C atoms increase the energy of the $\omega$ structure
compared with the ferromagnetic (FM) BCC.
The $\omega$ structure incorporating C atoms is also mechanically unstable
unless the C concentration is 25~at.\%.
It is concluded that
the $\omega$ structure is mostly unstable in steel,
and the $\omega$ structure in steel may be formed
under special atomic constraints at twin boundaries or other interfaces.
\end{abstract}


\begin{keyword}
First-principles calculations,
$\omega$ structure,
Steel
\end{keyword}

\end{frontmatter}


\section{Introduction}
\label{sec:introduction}

Nanometer-size domains of the $\omega$ structure have recently been reported in steel,
i.e., Fe-C-based alloys,
by detailed transmission electron microscopy observations
\cite{Ping2013, Liu2015}.
This motivated us to investigate the stability of the $\omega$ structure in steel.
The present authors' group has emphasized the importance of the $\omega$ structure
on transformation between the BCC and the FCC structures in metallic systems.
Togo and Tanaka
developed a search algorithm for transformation pathways
based on a systematic set of first-principles calculations
and revealed that
the $\omega$ structure was located on a transformation pathway
between the BCC and the FCC structures
\cite{Togo2013}.
Ikeda et al.
suggested that
the pressure-induced phase transition between the BCC and the FCC Fe at high temperature
occurred along this transformation pathway
\cite{Ikeda2014}.
The present authors recently performed
a systematic investigation into the $\omega$ structure of transition elements
\cite{Ikeda2016_omega_TE}.
The elemental $\omega$ Fe with antiparallel magnetic moments ($+--$ magnetic state)
was found to be the lowest in energy among the investigated magnetic states
and to be mechanically stable.
The $+--$ $\omega$ Fe was, however, 170~meV/atom higher in energy
than the ferromagnetic (FM) BCC Fe.
This implies that the elemental $\omega$ Fe should be difficult to be formed.
In experiments, however,
the nanometer-size $\omega$ structure was observed in steel
\cite{Ping2013, Liu2015}.
The largest difference between elemental Fe and steel may be the presence of C atoms,
but little has been known about effects of the C atoms on the stability
of the $\omega$ structure in steel.

Here we report a first-principles study on the stability of the $\omega$ structure in steel
with special interests in effects of interstitial C atoms.
Four different magnetic states of the $\omega$ structure investigated in our previous study
\cite{Ikeda2016_omega_TE}
(see Fig.~1(c) in Ref.~\cite{Ikeda2016_omega_TE} for notations of the magnetic states)
are focused on.
Possible interstitial sites for C atoms in the $\omega$ structure are
first systematically searched,
and then the energy of the $\omega$ structure is compared with that of the FM BCC
at several C concentrations using supercell models.
Finally,
mechanical stability of the $\omega$ structure is analyzed based on phonon frequencies
at the $\Gamma$ point of the supercell models.

\section{Computational details}


The $\omega$ structure belongs to the hexagonal crystal system,
and hence its primitive unit cell is specified by two lattice constants
$a_{\omega}$ and $c_{\omega}$.
The $\omega$ structure has three atoms inside the primitive unit cell,
and their positions are
$(0, 0, 0)$,
$(2/3, 1/3, 1/2)$,
and
$(1/3, 2/3, 1/2)$
in fractional coordinates.
The $\omega$ structure can actually be obtained from the BCC structure
by repeating to collapse a pair of neighboring $\{111$\} planes
and to hold the next plane unaltered.

Four magnetic states, which were focused on in our previous study
\cite{Ikeda2016_omega_TE},
were investigated for the $\omega$ structure.
The FM BCC structure and cementite Fe$_{3}$C$_{1}$ were also calculated for comparison.
The $\omega$-based BCC unit cell
(see Fig.~1(b) in Ref.~\cite{Ikeda2016_omega_TE} for details)
was used for the calculation of the FM BCC structure
so that computational conditions for the $\omega$ and the BCC structures are
as similar to each other as possible.


To systematically search possible interstitial sites for C atoms
in the $\omega$ structure in Fe,
we used the $\omega$ Fe$_{24}$C$_{1}$ model composed of the $2 \times 2 \times 2$ supercell
of the primitive $\omega$ unit cell for Fe and an interstitial C atom.
The possible interstitial sites were searched in the following procedure.
First, we divided a primitive $\omega$ unit cell into the $6 \times 6 \times 4$ mesh
and put a C atom on the mesh points that are symmetrically inequivalent to each other.
The numbers of the inequivalent points were 30 for the $++-$ magnetic state
and 21 for the other magnetic states.
Then,
we optimized lattice parameters and internal atomic positions of the structures
of the supercell models.
Some of the supercell models showed atomic positions largely deviated
from those of the initial $\omega$ structure after the structural optimization.
These models were excluded from further consideration.
Similarly, some models were excluded
because the initially given magnetic state was broken after the structural optimization.


Energies of the models were calculated based on the
plane-wave basis projector augmented wave method
\cite{Blochl1994}  
in the framework of density-functional theory
within the generalized gradient approximation of the Perdew-Burke-Ernzerhof form
\cite{Perdew1996}  
as implemented in the \textsc{VASP} code
\cite{
Kresse1995, 
Kresse1996, 
Kresse1999}.
A plane-wave energy cutoff of 400~eV was used.
%
The Brillouin zones were sampled by the $\Gamma$-centered
$12 \times 12 \times 18$ mesh per primitive $\omega$ unit cell,
and the Methfessel-Paxton scheme~\cite{Methfessel1989} with a smearing width of 0.4~eV
was employed.
Total energies were minimized until the energy convergences to be
less than 10$^{-8}$~eV.
Lattice parameters and internal atomic positions were optimized
under zero external stress.

Phonon frequencies were investigated within the harmonic approximation
for a lattice Hamiltonian using the finite-displacement method.
Atomic displacements of 0.01~{\AA}
were used to obtain the second-order force constants.
Phonon modes at the $\Gamma$ point of the supercell models were calculated
to investigate their mechanical stability.
When there are phonon modes with imaginary frequencies,
the structure is considered to be mechanically unstable.
The \textsc{PHONOPY} code~\cite{Togo2015, Togo2008} was used
for these phonon calculations.

\section{Results and discussion}
\label{sec:results}

In our previous study
\cite{Ikeda2016_omega_TE},
we made a systematic first-principles study
on thermodynamical and mechanical stability of the $\omega$ structure
in 27 transition elements
(Sc to Cu, Y to Ag, and Lu to Au).
Only the $\omega$ structures of the group 4 elements (Ti, Zr, and Hf),
the group 7 elements (Mn, Tc, and Re), and Fe were found to be mechanically stable
in their lowest-energy magnetic states.
For the elemental $\omega$ Fe,
the $+--$ magnetic state was the lowest in energy
among the investigated magnetic states.
It was, however, 170~meV/atom higher in energy than the FM BCC Fe.
This implies that
the elemental $\omega$ Fe is difficult to be formed from the viewpoint of
thermodynamical stability.
In the following,
we investigate whether the presence of interstitial C atoms stabilizes
the $\omega$ structure in Fe.


\begin{table}[btp]
\begin{center}
\caption{
Calculated energies of the $\omega$ Fe$_{24}$C$_{1}$ models in meV/(Fe~atom).
The energies are relative to that for the FM BCC Fe$_{24}$C$_{1}$ models
where the C atom is located at an octahedral site.
The first column shows the initial position of the interstitial C atom in fractional coordinates
for a primitive $\omega$ unit cell in the supercell models
before the structural optimization.
``NA'' indicates that the corresponding structure could not be obtained
because the optimized structure was largely deviated from the initial $\omega$ structure
and/or because the initially given magnetic state was broken
during the structural optimization.
\label{tb:energies_C_sites}
}
\vspace{2mm}
\footnotesize
\begin{tabular}{ccccc}
\hline
Initial position of C & FM  & $++-$ & $+--$ & NM   \\
\hline
$(\phantom{0/}0,\;           1/2,\; \phantom{0/}0)$   & 182 & 212   & 182   & 276  \\
$(1/6          ,\;           1/3,\; \phantom{0/}0)$   & 224 & NA    & 196   & 299  \\
$(1/6          ,\;           1/3,\; 1/2          )$   & 207 & NA    & 214   & 325  \\
$(\phantom{0/}0,\;           1/6,\; 1/2          )$   & 216 & NA    & 223   & 330  \\
$(\phantom{0/}0,\;           1/2,\; 1/2          )$   & 251 & NA    & 274   & 389  \\
$(1/3          ,\;           2/3,\; \phantom{0/}0)$   & NA  & 365   & NA    & 417  \\
$(\phantom{0/}0,\; \phantom{0/}0,\; 1/2          )$   & NA  & NA    & 349   & 428  \\
\hline
\end{tabular}
\end{center}
\end{table}

\begin{figure}[tbp]
\begin{center}
\includegraphics[width=\linewidth]{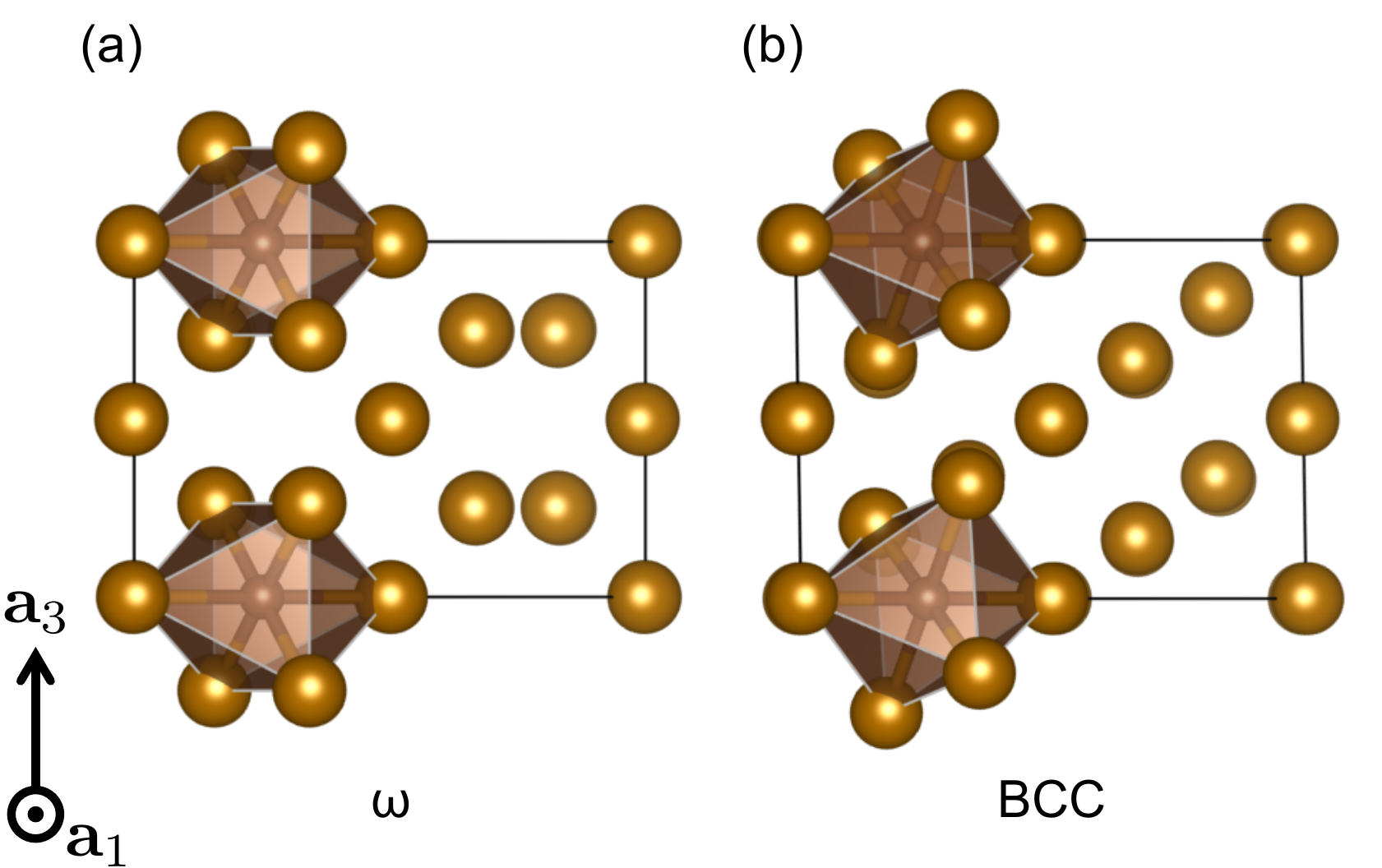}
\end{center}
\caption{
(Color online)
(a) Structure of the $\omega$ Fe$_{24}$C$_{1}$ model
where the C atom is located at the octahedral site.
(b) Structure of the BCC Fe$_{24}$C$_{1}$ model
where the C atom is located at the octahedral site.
Gold spheres represent Fe atoms,
and grey spheres inside the octahedra represent C atoms.
$\textbf{a}_{1}$ and $\textbf{a}_{3}$ denote two lattice vectors of the models.
Visualization is performed using the \textsc{VESTA} code
\cite{Momma2011}.
\label{fig:octahedral_sites}
}
\end{figure}

First,
we search possible interstitial sites for C atoms in the $\omega$ structure
based on the systematic procedure described above.
Table~\ref{tb:energies_C_sites} summarizes
calculated energies of the obtained $\omega$ Fe$_{24}$C$_{1}$ models.
Calculations from the inequivalent initial positions of the C atom
converged to seven positions as listed in Table~\ref{tb:energies_C_sites}.
When the initial position of the C atom is $(0, 1/2, 0)$ in fractional coordinates
for a primitive $\omega$ unit cell,
the energy is the lowest among those of the interstitial sites for all the magnetic states.
The $(0, 1/2, 0)$ interstitial site can be referred to as the octahedral site
of the $\omega$ structure,
as shown in Fig.~\ref{fig:octahedral_sites}(a).
The octahedral site of the $\omega$ structure is actually similar to
that of the BCC structure, as shown in Fig.~\ref{fig:octahedral_sites}(b).
It has been well investigated that
C atoms favor the octahedral site in the BCC Fe
\cite{Jiang2003}.  
The present calculations reveal
good correspondence of the favorable interstitial sites
between the two crystal structures.



\begin{figure}[tbp]
\begin{center}
\includegraphics[width=\linewidth]{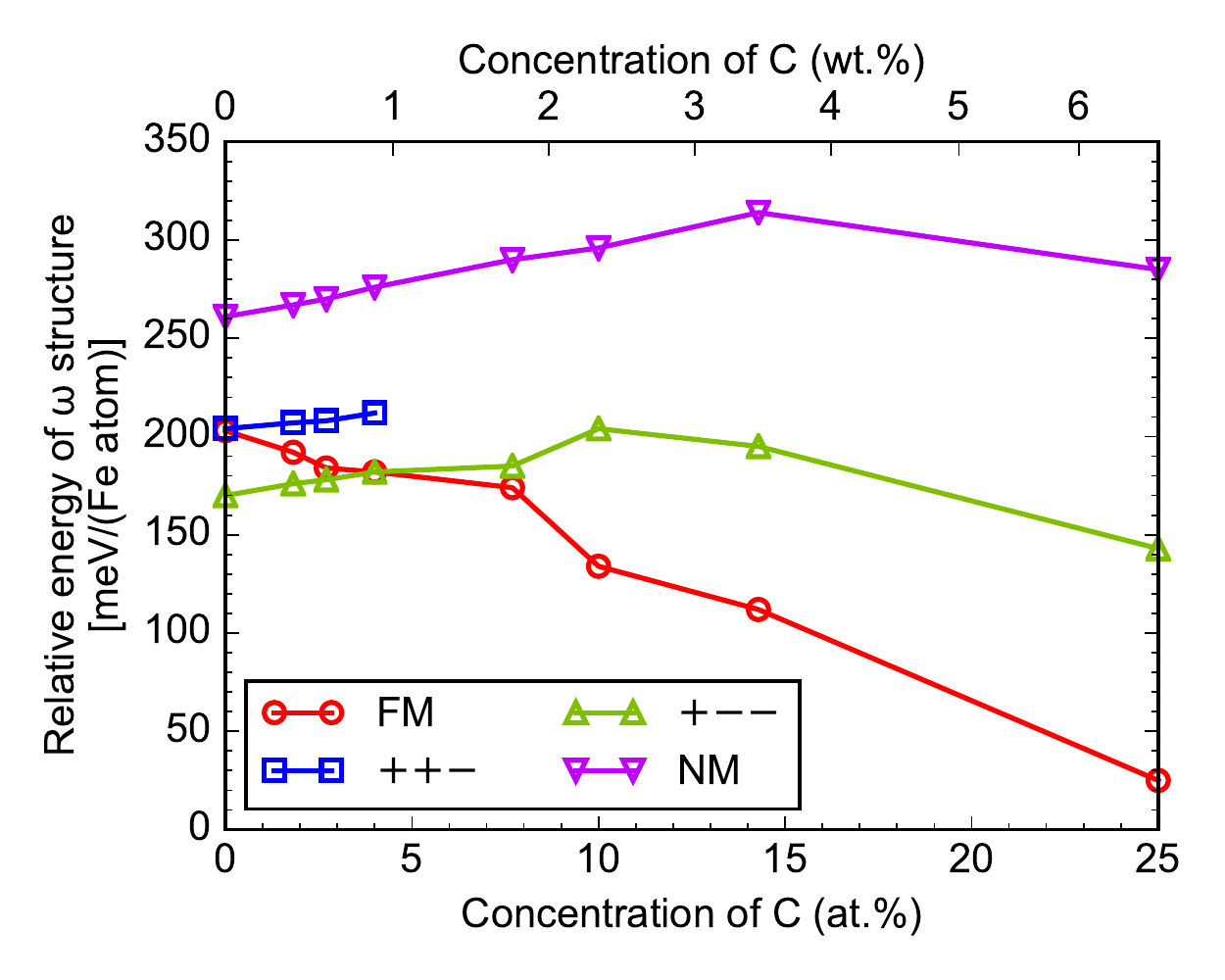}
\end{center}
\caption{
(Color online)
Calculated energies of the $\omega$ Fe$_{n}$C$_{1}$ models where the C atom is located
at an octahedral site
with respect to the concentration of C.
The energies are relative to that of the FM BCC Fe$_{n}$C$_{1}$ models
where the C atom is located at an octahedral site.
Lines are guides for the eyes.
Note that the energy of the $++-$ $\omega$ structure cannot be obtained
above 4~at.\%~C
because magnetic states of the optimized structures were broken
during the structural optimization.
\label{fig:energies_supercells}
}
\end{figure}

To investigate dependence of energies on the concentration of C atoms,
we use the
$1 \times 1 \times 1$,
$1 \times 1 \times 2$,
$1 \times 1 \times 3$,
$2 \times 2 \times 1$,
$2 \times 2 \times 2$,
$2 \times 2 \times 3$,
and
$3 \times 3 \times 2$
supercells of the primitive $\omega$ unit cell
with a C atom at an octahedral site.
Hereafter supercell models incorporating a C atom
are collectively referred to as Fe$_{n}$C$_{1}$
($n = 3, 6, 9, 12, 24, 36, 54$),
where $n$ indicates the number of Fe atoms in the supercell models.
Energies of the FM BCC Fe$_{n}$C$_{1}$ where the C atom is located at an octahedral site
are also investigated using the corresponding supercells for comparison.
Figure~\ref{fig:energies_supercells} shows the calculated energies
of the $\omega$ Fe$_{n}$C$_{1}$ relative to those of the FM BCC Fe$_{n}$C$_{1}$.
The $\omega$ structure is higher in energy than the FM BCC
up to 25~at.\%~C for all the magnetic states.
This result indicates that the $\omega$ structure is thermodynamically less favorable
than the FM BCC even when interstitial C atoms are incorporated.
Furthermore,
the relative energy of the $\omega$ structure below 14~at.\%~C
is higher than that of the elemental $\omega$ Fe
for the $++-$, $+--$, and nonmagnetic (NM) states.
This indicates that
interstitial C atoms thermodynamically destabilize the $\omega$ structure
in the $++-$, $+--$, and NM states at these C concentrations compared with the FM BCC.
For the FM $\omega$ structure, in contrast,
the relative energy monotonously decreases as the C concentration increases.
While the FM $\omega$ structure is 32~meV/(Fe~atom) higher in energy than the $+--$ $\omega$
when no interstitial C atoms are incorporated,
the energy difference becomes smaller as the C concentration increases.
Finally the FM $\omega$ structure becomes lower in energy than the $+--$ $\omega$
above 4~at.\%~C.


\begin{table}[tbp]
\begin{center}
\caption{
Numbers of imaginary phonon modes at the $\Gamma$ point
of the FM and $+--$ $\omega$ Fe$_{n}$C$_{1}$
where the C atom is located at an octahedral site.
\label{tb:imaginary_modes_concentration}
}
\vspace{2mm}
\footnotesize
\begin{tabular}{cccc}
\hline
 & Supercell size & FM & $+--$ \\
\hline
Fe$_{  3}$C$_{1}$ & $1 \times 1 \times 1$ &  0 & 0 \\
Fe$_{  6}$C$_{1}$ & $1 \times 1 \times 2$ &  1 & 1 \\
Fe$_{  9}$C$_{1}$ & $1 \times 1 \times 3$ &  2 & 1 \\
Fe$_{ 12}$C$_{1}$ & $2 \times 2 \times 1$ &  6 & 3 \\
Fe$_{ 24}$C$_{1}$ & $2 \times 2 \times 2$ &  7 & 2 \\
Fe$_{ 36}$C$_{1}$ & $2 \times 2 \times 3$ &  8 & 2 \\
Fe$_{ 54}$C$_{1}$ & $3 \times 3 \times 2$ & 19 & 3 \\
Fe                &                       &  1 & 0 \\
\hline
\end{tabular}
\end{center}
\end{table}

Table~\ref{tb:imaginary_modes_concentration} summarizes
the number of imaginary phonon modes at the $\Gamma$ point
of the FM and $+--$ $\omega$ Fe$_{n}$C$_{1}$
where the C atom is located at an octahedral site.
Except for the $\omega$ Fe$_{3}$C$_{1}$,
which corresponds to 25~at.\%~C,
all the models with interstitial C atoms have
one or more imaginary phonon modes at the $\Gamma$ point
and hence are mechanically unstable.
Although the elemental $+--$ $\omega$ Fe is mechanically stable
\cite{Ikeda2016_omega_TE},
it becomes mechanically unstable
once it incorporates interstitial C atoms
unless the C concentration is 25~at.\%.
This result indicates that
the interstitial C atoms mechanically destabilize the $\omega$ structure in Fe.
Although the $\omega$ Fe$_{3}$C$_{1}$ shows no imaginary phonon modes at the $\Gamma$ point,
the FM $\omega$ Fe$_{3}$C$_{1}$ is still 25~meV/(Fe~atom) higher in energy
than the FM BCC Fe$_{3}$C$_{1}$ and hence is thermodynamically less favorable.
Furthermore,
it is well known that cementite Fe$_{3}$C$_{1}$ is thermodynamically much more favorable
in such a high C concentration region.
Our first-principles calculations actually show that
the FM cementite Fe$_{3}$C$_{1}$ is 111 and 135~meV/(Fe~atom) lower in energy
than the FM BCC Fe$_{3}$C$_{1}$ and the FM $\omega$ Fe$_{3}$C$_{1}$, respectively.


The present calculations demonstrate that
the $\omega$ structure is thermodynamically less favorable than the FM BCC
irrespective of the C concentration up to 25~at.\%.
Furthermore,
the FM and the $+--$ $\omega$ structures are mechanically unstable
once they incorporate C atoms
unless the C concentration is 25~at.\%.
These results imply the instability of the $\omega$ structure in Fe-C alloys.
We conclude that the $\omega$ structure is mostly unstable
and cannot exist even as a metastable state in Fe-C alloys.
In experimental reports
\cite{Ping2013, Liu2015},
the $\omega$ structure was observed in steel
as nanometer-size domains in the BCC structure
or as nanometer-thick layers at twin boundaries of the BCC.
Based on the present theoretical results,
we suggest that the $\omega$ structure in steel is formed
under special atomic constraints at twin boundaries or other interfaces.
It is actually known that twin boundaries in metallic systems
sometimes show special structures such as the $9R$ structure in Cu and Ag
\cite{Wolf1992, Ernst1992}.
Since the $\omega$ structure acts as the transition state of a transition pathway
between the BCC and the FCC structures,
the $\omega$ structure in steel may be a kind of residue of the martensitic transformation.
Another possible factor may be enrichment of solute elements other than interstitial C atoms
in the nanometer-scale regions.
The specimens used in the experiments actually contain Si, Mn, and/or Cr as well as C
\cite{Ping2013, Liu2015}.


Finally we comment on theoretical calculations in Ref.~\cite{Ping2013}.
Two points are inconsistent with the present results.
First,
the energy of the $\omega$ structure relative to that of the BCC structure
is substantially different at 7.7 and 14.3~at.\%~C
between Ref.~\cite{Ping2013} and the present report.
The authors in Ref.~\cite{Ping2013} reported that
the relative energies are 142 and 78~meV/(Fe~atom) at 7.7 and 14.3~at.\%~C, respectively.
In contrast,
the present calculations show that
the energies of the FM $\omega$ structure relative to those of the FM BCC are
174 and 112~meV/(Fe~atom) at 7.7 and 14.3~at.\%~C, respectively,
which are much larger than the values in Ref.~\cite{Ping2013}.
Since the authors in Ref.~\cite{Ping2013} do not refer to magnetic states
of the $\omega$ structure incorporating C atoms,
it is difficult to discuss possible reasons for the discrepancies.
Note that the present calculations demonstrate that
the FM $\omega$ structure is the lowest in energy among the magnetic states
above 4~at.\%~C,
which was not reported in Ref.~\cite{Ping2013}.
Second,
the authors of Ref.~\cite{Ping2013} claimed that
the BCC Fe$_{3}$C$_{1}$ changed to the $\omega$ Fe$_{3}$C$_{1}$
without energy barriers.
They did not, however, show the computational details for this issue.
Contrary to their results,
our FM BCC Fe$_{3}$C$_{1}$ is mechanically stable and is 25~meV/(Fe~atom) lower in energy
than the FM $\omega$ Fe$_{3}$C$_{1}$.

\section{Conclusion}

Stability of the $\omega$ structure in steel is investigated based on first-principles
with special interests in effects of interstitial C atoms.
The energy of the $\omega$ structure is compared with that of the FM BCC,
and mechanical stability is analyzed based on phonon frequencies
at the $\Gamma$ point of supercell models.

Possible interstitial sites for C atoms in the $\omega$ structure are systematically searched,
and the octahedral site is found to be the most favorable among the interstitial sites.
The $+--$ $\omega$ Fe$_{n}$C$_{1}$ where the C atom is located at an octahedral site is
the lowest in energy among the magnetic states below 4~at.\% C,
while the FM $\omega$ Fe$_{n}$C$_{1}$ is the lowest above this concentration.

Even when C atoms are incorporated in the $\omega$ structure,
it is thermodynamically less favorable than the FM BCC.
Furthermore,
the FM and $+--$ $\omega$ structures are
mechanically unstable once they incorporate C atoms
unless the C concentration is 25~at.\%.
These results indicate that
interstitial C atoms destabilize the $\omega$ structure in Fe-C alloys.
It is concluded that the $\omega$ structure is mostly unstable and cannot exist
even as a metastable state in Fe-C alloys.
The $\omega$ structure in steel observed in experiments may be stabilized
under special atomic constraints at twin boundaries or other interfaces.


\section*{Acknowledgments}

The authors thank D. Ping for providing valuable information on the observation
of the $\omega$ structure in steel.
Funding by the Ministry of Education, Culture, Sports,
Science and Technology (MEXT), Japan, through Elements Strategy Initiative for
Structural Materials (ESISM) of Kyoto University,
is gratefully acknowledged.

 
\section*{\refname} 

\end{document}